\documentclass[twocolumn,showpacs,prb]{revtex4}% Physical Review B
\usepackage{graphicx}
\usepackage{bbold}
\usepackage{amsmath}
\usepackage{epsfig}
\usepackage{subfigure}
\usepackage{sidecap}

\newcommand{\bhy}{\hat{\bf y}}

\newcommand{\eqa}{\begin{eqnarray}}
\newcommand{\eqe}{\end{eqnarray}}

\newcommand{\bv}{{\bf v}}
\newcommand{\bmu}{{\boldsymbol\mu}}
\newcommand{\bxi}{{\boldsymbol\xi}}

\newcommand{\bh}{{\bf h}}
\newcommand{\bu}{{\bf  u}}
\newcommand{\bx}{{\bf x}}
\newcommand{\bff}{{\bf  f}}

\newcommand{\bk}{{\bf k}}

\newcommand{\br}{{\bf r}}

\newcommand{\bQ}{{\bf Q}}
\newcommand{\bA}{{\bf A}}
\newcommand{\hg}{\hat{\cal G}^{-1}}

\begin{document}

\title{Vortex Wall Dynamics and Pinning in Helical Magnets}

\author{Bahman Roostaei$^{1,2}$}

\affiliation{$^1$ Institut f\"ur Theoretische Physik, Universit\"at zu K\"oln, Z\"ulpicher Str. 77, D-50937 K\"oln, Germany}
\affiliation {$^2$ Department of Physics, Indiana University-Purdue University Indianapolis, Indiana 46202, USA}

\date{\today}
\begin{abstract}
Domain walls formed by one dimensional array of vortex lines have been recently predicted to exist in disordered helical magnets and multiferroics. These systems are on one hand analogues to the vortex line lattices in type-II superconductors while on the other hand they propagate in the magnetic medium as a domain boundary. Using a long wavelength approach supported by numerical optimization we lay out detailed theory for dynamics and structure of such topological fluctuations at zero temperature in presence of weak disorder. We show the interaction between vortex lines is weak. This is the direct consequence of the screening of the vorticity by helical background in the system. We explain how one can use this result to understand the elasticity of the wall with a vicinal surface approach. Also we show the internal degree of freedom of this array leads to the enhancement of its mobility. We present estimates for the interaction and mobility enhancements using the microscopic parameters of the system. Finally we determine the range of velocities/force densities in which the internal movement of the vortex wall can be effective in its dynamics.
\end{abstract}
{}

\maketitle

\section{Introduction}
\label{sec0}
%\begin{figure}[t]
%\includegraphics[width=8.5cm]{3D_vortex.eps}
%\caption{Schematic illustration of the cross section of a vortex line (oval region) formed between two anti-chiral domains.}
%\label{3D_vortex}
%\end{figure}
Domain walls and domain patterns in magnetic materials to a large extent determine the controllability of the
magnetic behavior of devices, a crucial factor in technology of today's spintronics and magnetic storage systems. Amongst various types of magnetic
structures, helical magnets are in current interest demanding detailed studies of their properties under various circumstances. With the discovery of multiferroics\cite{Cheong2007}, materials such as RMnO$_{3}$ in which R $\in \{$Y,Tb,Dy,Ho$\}$\cite{8,9,10,11,14} with coupled magnetic and electric properties and their intrinsic spiral magnetic order the necessity of such studies is understandable more than ever. A new class of topological domain walls has been predicted to exist in Helical magnets in recent studies\cite{Li2012}. These domain walls separate two domains with opposite chirality and posses vorticity. These domains are indeed observed in circularly polarized X-ray scattering experiments\cite{Lang2004}. Static domain patterns in these observations suggest strong pinning of the walls. In another recent study\cite{Nattermann} the pinning mechanism of domain walls in these systems has been presented. Following this study, with an emphasis on pure vortex walls we present a detailed theory for structure and dynamics of these walls. We show that inside the topologically protected vortex walls\cite{Li2012} the interaction between vortices is weak. We explain the effect of disorder on their structure and dynamics. We also explain how the internal dynamics of such walls can affect their mobility in presence of weak disorder. This is indeed a new realization of previously considered systems\cite{Parkin2008,Lecomte2009} where internal degree of freedom enhances the mobility of the magnetic structure and therefore reduces the threshold force density.

 We consider a major class of bulk three dimensional helical magnets, the centrosymmetric (CS) systems in which the magnetic moment distribution is invariant
under space inversion. Examples of such materials are rare earth metals such as Ho,Tb,Dy\cite{Jensen}. At low enough temperature spins lie on basal ferromagnetic planes which we take to be y-z plaes. The helical ground state in such materials originates from RKKY exchange interaction between local spins in basal planes. This interaction effectively results in competing nearest neighbor ferromagnetic ($J<0$) and next nearest neighbor antiferromagnetic coupling ($J'>0$) along the x-axis perpendicular to basal planes. The local spins thus lower their energy at zero temperature by ordering in a spiral state in which all spins in each plane makes an angle $\theta=\arccos(|J/4J'|)$ with respect to its nearest neighbor plane. Here throughout this paper we choose this axis along $\hat{\bf x}$. We assume isotropy in the nearest neighbor ferromagnetic interaction. With these assumption the Ginzburg-Landau hamiltonian density for the local spins in helical rare earth CS magnets frequently used in literature can be written in the following form:
\eqa\label{GL-CS}
h[{\bf m}]={J\over a}\left[{a^2\over 4}\left(\partial_x^2{\bf m}+q^2 {\bf m}\right)^2+(\nabla_\perp{\bf
m})^2\right]
\eqe
in which $a$ is the distance between the lattice points and $q=\theta/a$ is the magnitude of helix wavevector. Here after we use the notation $\perp$ to denote the coordinates $\{y,z\}$. This Hamiltonian has been obtained from proper long wavelength expansion of the discrete classical Heisenberg model defined on a cubic lattice:
\begin{equation}\label{discrete}
H = J\sum_n\sum_{\langle i,j\rangle} {\bf m}_{i,n}\cdot {\bf m}_{j,n+1} + J'\sum_{i,n} {\bf m}_{i,n}\cdot {\bf  m}_{i,n+2}
\end{equation}
in which $i,j$ specifies the spins in y-z plane and $n$ specifies each plane. Here throughout this paper we ignore the small out of plane component of the spins ($m_x$) and assume ${\bf m}\approx\{0,\cos\varphi,\sin\varphi\}$. As a result the long wavelength Hamiltonian density becomes:
\eqa\label{LWV}
h={J\over a}\left\{(\partial_\perp\varphi)^2+{a^2\over
4}\left[(\partial_x\varphi)^2-q^2\right]^2+{a^2\over 4}(\partial_x^2\varphi)^2\right\}
\eqe
Here the second derivative terms has the role of stabilizing the helix structure along the x-axis. The ground state is clearly reached by $(\partial_x\varphi)^2=q^2$. This state is degenerate with respect to the chirality (sense of rotation) of the helix.
\section{vortex wall}\label{vortex wall}
%%%%%%%%%%%%%%%%%%%%%%%%%%%%%%%%%% ansatz verification %%%%%%%%%%%%%%%%%%%%%%%%%%%%%%%
\begin{figure}[t]
\includegraphics[width=8.5cm]{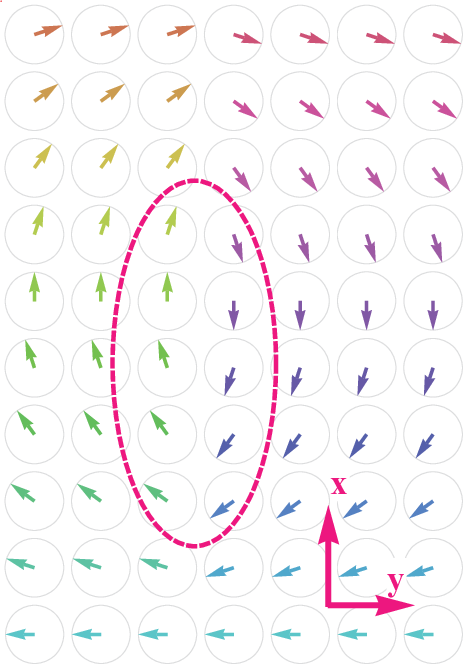}
\caption{Illustration of vorticity. For better visualization spins have been represented to lie in the x-y plane \textit{parallel} to the helical axis while in the text the spins are assumed to lie in y-z planes perpendicular to the helical axis. The dashed line shows the core of the vortex. The line integral around such path or any other path around this core yields non-zero values (see text).}
\label{path_integral}
\end{figure}
%%%%%%%%%%%%%%%%%%%%%%%%%%%%%%%%%%%%%%%%%%%%%%%%%%%%%%%%%%%%%%%%%%%%%%%%%%%%%%%%%%%%%%%%%%%%%%%
\begin{figure}
\begin{center}
\subfigure[]{
\includegraphics[scale=.8]{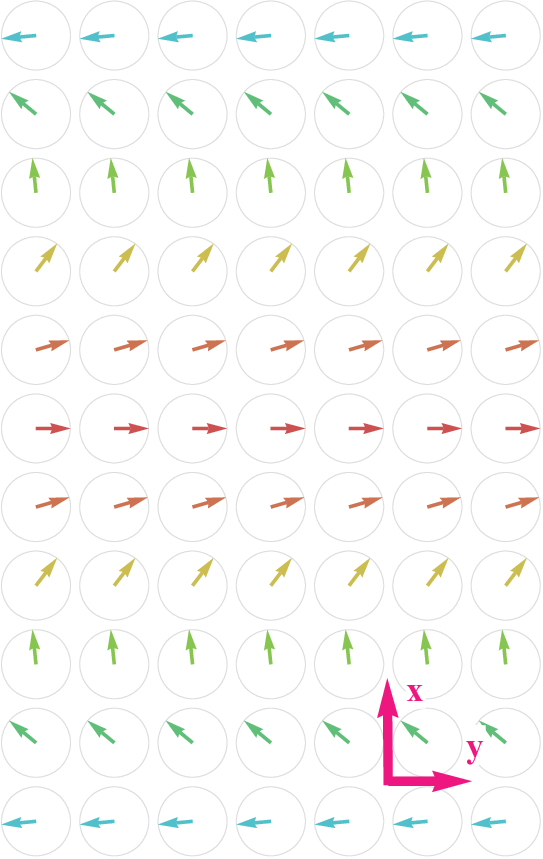}}
\subfigure[]{
\includegraphics[scale=.6]{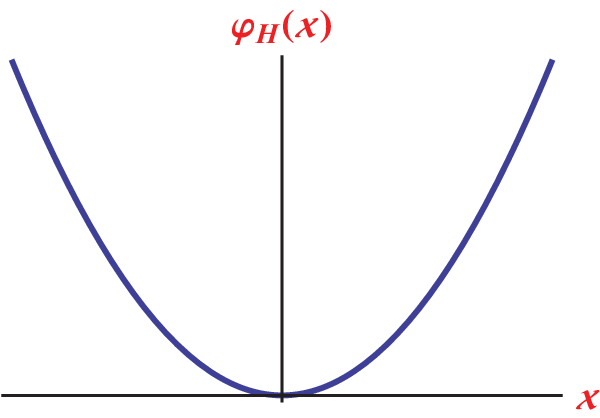}}
\end{center}
\caption{(a) Illustration of Hubert wall. For better visualization spins have been represented to lie in x-y plane \textit{parallel} to the helical axis. (b) Hubert wall profile function. The chirality changes sign at the origin.}
\label{hubert}
\end{figure}
%%%%%%%%%%%%%%%%%%%%%%%%%%%%%%%%%%%%%%%%%%%%%%%%%%%%%%%%%%%%%%%%%%%%%%%%%%%%%%%%%%

As was mentioned in the introduction the ground state of the helical system is degenerate with respect to chirality and the domains with opposite chirality have been observed in polarized X-ray scattering experiments. Li, Nattermann and Pokrovsky indicated\cite{Li2012} that the domain wall between two spirals with opposite chirality side by side (their propagation vectors being parallel) must posses vorticity. This can be easily seen by taking a line integral over a closed path that passes along $N_v/2$ pitches of helix on one side of the wall and backwards on the other side (see Fig. \ref{path_integral}). Over such path:
\eqa
\oint \nabla\varphi\cdot d{\bf l} = 2\pi N_v.
\eqe
in which $N_v$ is half the number of pitches hence the number of vortices. For a single vortex line in this case one needs to solve the nonlinear saddle point differential equation associated with the
hamiltonian (\ref{LWV}) along with the boundary condition $\nabla\times\nabla\varphi=2\pi\delta^2(\br)\hat{\bf z}$. Li \text{et. al.}\cite{Li2012} have used variational method using an ansatz and separation of scales to find the solution. In the next section we show that their result is
also confirmed with numerical minimization. The variational
approximation that is used for different scales is a function of the form $\varphi=\tan^{-1}(\lambda y/x)$ which
minimizes the energy by varying $\lambda$ however done only for a certain distance $r$ defined
as: $r^2=x^2+\lambda(r)^2y^2$. The result is\cite{Li2012}:
\eqa\label{vortex energy analytic}
\epsilon_v(r)={\pi J\over a^3}\log^{1/2}(r/a)\left[{5\over 64}+\theta^2\log(r/a)\right]^{1/2}
\eqe
and:
\eqa
\lambda^2(r)=\theta^2+{5/64\over\log(r/a)}
\eqe
For large enough $r\gg a e^{0.08/\theta^2}$ (the constant 0.08 in the exponential is not important since it comes from a variational calculation) the logarithmic term in the bracket dominates and the energy behaves logarithmic similar to conventional vortices in two dimensional XY models. For scales much smaller than this, close to the core of the vortex the value of the energy behaves as square root of logarithm. This behavior can be easily understood from the anisotropic saddle point equation associated with the Hamiltonian (\ref{LWV}):
\eqa\label{saddle}
\left\{4\nabla_\perp^2+a^2\left[6(\partial_x\varphi)^2-2q^2-\partial_x^2\right]\partial_x^2\right\}\varphi=0.
\eqe
At larger length scales and away from core $(\partial_x\varphi)^2\approx q^2$ which approximates the above with a laplace equation. The vortex solutions to laplace equation is well known to diverge logarithmically with linear system size. Closer to the core the energy of the vortex line in the helical system on the other hand behaves differently as can be seen in (\ref{vortex energy analytic}) which is square root of logarithm behavior.
As a result this type of vortices in helical systems posses a layered structure.

 For a vortex wall consisting of an array of $2N_v+1\gg 1$ vortex lines we can introduce the following ansatz:
\eqa\label{linear-comb}
\varphi(x,y)=\sum_{n=-N_v}^{N_v} \tan^{-1}{\lambda y\over x-x_n}
\eqe
in which $x_n=n\pi/q$. This solution indeed
satisfies the boundary conditions $\varphi(x,y)=\pm qx$ and $\partial_y\varphi=0$ as $y\rightarrow
\pm\infty$. This can be seen by noting the following identities for $N_v\gg 1$:
\eqa
\partial_x\varphi &=& \sum_{n=-N_v}^{N_v} {-\lambda y\over (x-x_n)^2+\lambda^2 y^2}\nonumber\\
&=& {iq\over 2}\left(\cot \bar{w} -\cot w\right)\\
\partial_y\varphi &=& \sum_{n=-N_v}^{N_v} {\lambda (x-x_n)\over (x-x_n)^2+\lambda^2 y^2}\nonumber\\
&=& {q\lambda\over 2}\left(\cot \bar{w}+\cot w\right)
\eqe
in which $w=q(x+i\lambda y)$. Clearly the above two expressions satisfy the boundary conditions.

From the above calculation it is interesting to observe that the field distribution close to the
vortex wall in this ansatz is identical to the field of a single vortex:
\eqa\label{close to core}
\partial_x\varphi(w\rightarrow 0)&=& {iq\over 2}\left({1\over \bar{w}}-{1\over w}\right)={-\lambda
y\over x^2+\lambda^2 y^2}\nonumber\\
\partial_y\varphi(w\rightarrow 0)&=& {\lambda q\over 2}\left({1\over \bar{w}}+{1\over
w}\right)={\lambda
x\over x^2+\lambda^2 y^2}\nonumber\\
\eqe
indicating that the value of the variational parameter in smaller scales will be identical to the value obtained for a single vortex
but this would not be the case for larger length scales (see below). Moreover this is indicating the fact that the individual vortices almost do not disturb each other in this variational solution, hence their interaction must be weak. This will be confirmed also numerically in the next section. It is very instructive to derive an analytic expression for the deformation energy of
such vortex lattice as we proceed. The goal of this part will be to have a fundamental understanding of the interaction between vortex lines in helical structures.

We start with the Hamiltonian (\ref{LWV}). In presence of the vortex wall at larger length scales we can write the effect of the wall deformations in the field as:
\eqa
\partial_x\varphi = A_x(\br)+\psi
\eqe
in which $A_x(\br)=q\text{sgn}_q[y-u_y(\bx_\parallel)]$ is a smooth sign function representing the helical structure on both sides of the wall and varies on
the scales of $q^{-1}$, $u_y(\bx_\parallel)$ is the position of the wall at each point of its
plane ${\bf x}_\parallel=\{x,z\}$ and $\psi$ is the remaining of the effects which is supposedly small. Up to
second order in $\psi$ we can write $\left[(\partial_x\varphi)^2-q^2\right]^2\approx 4q^2 \psi^2$.
However we re-insert the full $\varphi$ field into the Hamiltonian in order to preserve the
non-perturbative effect of the vortices in the field:
\eqa
h &\approx& {J\over
a}\left\{(\partial_\perp\varphi)^2+\theta^2\left[\partial_x\varphi-A_x(\br)\right]^2\right\}.
\eqe
Here we have ignored the highest derivative of $\varphi$ assuming $\partial_x\psi\ll q\psi$ which is valid at our long wavelength approximation.
Introducing the rescaled coordinates $\tilde{\br}=\{x/\theta,y,z\}$ and fields
$\tilde{\bA}=\theta\{A_x,0,0\}$ we can write the above energy in the form of the vortex line
lattice energy in a type II superconductor:
\eqa
h \approx  {J\over
a}\left[\tilde{\nabla}\varphi-\tilde{\bA}(\tilde{\br})\right]^2.
\eqe
Here $\tilde{\nabla}=\{\partial_{\tilde{x}},\partial_y,\partial_z\}$. Note that by definition, for $y\gg u_y$ we have $\bA \approx + q \hat{\bx}$ and vice versa. In a superconducting analogy the vector $\bA$ plays the role of \textit{vector potential} which in reality represents the helical structure in the background. In the same analogy rotation of this vector potential represents a magnetic field ${\bf B}=\nabla\times\bA $. In the case of a flat wall $(u_y=0)$ this magnetic field will be ${\bf B}\approx 2q\delta_q(y)\hat{\bf z}$ in which $\delta_q(y)$ is a smooth dirac delta function over the scale $q^{-1}$. This magnetic field in the superconducting picture generates vortex lines along the $\hat{\bf z}$ direction in the $\varphi$ field. In reality \textit{vortex wall in helical magnets is the result of the rotation of the helical structure} represented here by the vector $\bA$.

 The saddle point equation from the above energy:
\eqa
\tilde{\nabla}\cdot \left[\tilde{\nabla}\varphi-\tilde{\bA}\right]=0
\eqe
must be solved in presence of the vortices:
\eqa\label{vortex-bounday}
\tilde{\nabla}\times\tilde{\nabla}\varphi={\bf J}(\tilde{\br})
\eqe
in which the vortex line density is defined as:
\eqa
{\bf J}(\br)=2\pi\sum_{n=-N_v}^{N_v}\int d{\br_n}\delta^3(\br-\br_n)
\eqe
with ${\bf r}_n(z)$ being the position of each point on the vortex line number $n$. The solution to the above system can be obtained by another vector potential ${\bf a}$ similar to normal procedure in solving classical magnetostatics problems:
\eqa
\tilde{\nabla}\varphi=\tilde{\bf A}+\tilde{\nabla}\times \bf a
\eqe
with a gauge freedom which we use by choosing $\tilde{\nabla}\cdot{\bf a}=0$. Inserting this
solution into (\ref{vortex-bounday}) we obtain the equation:
\eqa
\tilde{\nabla}^2{\bf a}=\tilde{\nabla}\times\tilde{\bf A}-\bf J
\eqe
which has the solution:
\eqa
{\bf a}(\tilde{\br})=\int_{\tilde{\br}'}
G(\tilde{\br}-\tilde{\br}')[\tilde{\nabla}'\times\tilde{\bf A}-{\bf J}(\tilde{\br}')]
\eqe
in which $G(\br)=(4\pi)^{-1}/|\br|$ is the Green's function for Laplace equation in three dimensions. Note that here the potential ${\bf a}$ is presenting the extra twist in magnetic configuration due to the vortices. On the other hand we already realized that the rotation of the helical structure at long wavelength approximation $\nabla\times\bA$ generates vortices. This indicates the fact that the vector potential $\bf a$ is the effect of the twist at the area where $\nabla\times \bA\neq {\bf J}$ meaning at the core of each vortex.  Now inserting $\bf a$ into the energy we obtain:
\eqa\label{helical background}
{\cal H}=J{\theta\over a}\int_{\tilde{\br},\tilde{\br}'}
G_{\tilde{\br}\tilde{\br}'}[\tilde{\nabla}\times\tilde{\bf A}-{\bf J}(\tilde{\br})]\cdot
[\tilde{\nabla}'\times\tilde{\bf A}-{\bf J}(\tilde{\br}')]
\eqe
Let's calculate the vector ${\bf D}=\nabla\times \bA-{\bf J}$ more explicitly. In order to do
that we use the large scale approximation in which the density of the vortex lines is uniform :
\eqa
{\bf D}(\br)={2\theta^2\over a}(\partial_z u_y\hat{\bf y}+\hat{\bf
z})[\delta_q(y-u_y)-\delta(y-u_y)]
\eqe
in which $\delta_q(y)$ comes from derivative of the sign function and is smooth version of delta function on a range $q^{-1}$. The function ${\bf D}$ is obviously non-zero only at a narrow region with the thickness $q^{-1}$ of the
order of the core of the vortices. The equation (\ref{helical background}) clearly shows
that the helical background represented by the vector potential $\bA$ screens the vorticity fields in large scales consistently with
our choice of the ansatz (\ref{linear-comb}). This screening results in weak interaction between the vortex lines in the domain wall.
This fact can be used as a basis of a vicinal surface model\cite{Nattermann} for the deformation of the vortex wall. We will briefly explain this model in later sections to justify our use of the quadratic elastic energy for studying the dynamics of the vortex walls.

%**********************************
\section{Hubert walls}
Hubert studied domain walls in helical magnets that are perpendicular to the helical axis\cite{Hubert} (here in y-z plane).
The sense of rotation gradually changes as one goes along the x-axis from one side of the Hubert wall to the other side parallel to the helical axis. The profile of the Hubert wall in continuum limit is obviously one
dimensional and is obtained by solving the saddle point equation (\ref{saddle}) with the boundary condition
$\varphi_H(x)=\pm q x $ as $x\rightarrow \pm\infty$. Luckily this equation has an exact solution
$\varphi_H(x)=\log[\cosh(qx)]$\cite{Fairbirn}. Fig. (\ref{hubert}) shows the behavior of this function. The slope of the curve indicates the chirality of the magnetization. As one moves from negative $x$ to positive $x$ the slope changes sign. This solution will lead to an energy per unit area $\sigma_H=2Jaq^3/3$ for a flat Hubert wall. We will use this value to calculate the deformations of the vortex walls in later sections.

%******************************************************
%Here we adopt the same method. First of all because we are only interested in long range
%$(>q^{-1})$ behavior of the domain wall elasticity we linearize the saddle point equation:
%\eqa
%\left[\theta^2\partial_x^2-{a^2\over
%4}\partial_x^4+\partial_\perp^2\right]\varphi=\theta^2\partial_x A(\br)
%\eqe
%The solution to the above linear differential equation can be written in the following general
%form:
%\eqa
%\varphi(\br)=\int_0^x A(x')dx'+\int_\bk e^{i\bk\cdot
%\bx_\parallel-Q_\pm(\bk)|x|/\theta}\alpha^{\pm}_\bk
%\eqe
%in which $Q_\pm(\bk)=\left|2q^2\pm2\sqrt{q^4-\bk^2/a^2}\right|^{1/2}$ and $\alpha_\bk$ is
%determined with the boundary condition $\varphi(u(\bx_\parallel),\bx_\parallel)=C$ with $C$ being
%an angular value uniform throughout the wall. For large scales $\bk\rightarrow 0$ we see that
%$Q_+(\bk)\approx 2q+{\cal O}(k^2)$ and $Q_-(\bk)\approx |\bk|/2\theta+{\cal O}(k^3)$.

\section{Numerical Investigation}\label{numerics}
%%%%%%%%%%%%%%%%%%%%%%%%%%%%%%%%%% ansatz verification %%%%%%%%%%%%%%%%%%%%%%%%%%%%%%%
%\begin{figure}[t]
%\includegraphics[width=6.5cm]{vortex_wall_picture.eps}
%\caption{Vector field illustration of numerically optimized interface between two
%ani-chiral phases. Here $\theta=0.32$.}
%\label{vortex_wall_picture}
%\end{figure}
%%%%%%%%%%%%%%%%%%%%%%%%%%%%%%%%%%%%%%%%%%%%%%%%%%%%%%%%%%%%%%%%%%%%%%%%%%%%%%%%%%%%%%%%%%%%%%
%\begin{figure}[t]
%\begin{minipage}[t]{0.45\linewidth}
%\includegraphics[width=8cm]{vortex_wall_picture.eps}
%\end{minipage}
%\hspace{0.5cm}
%\begin{minipage}[t]{0.45\linewidth}
%\includegraphics[width=8cm]{staircase.eps}
%\end{minipage}
%\caption{Left: Vector field illustration of numerically optimized interface between two
%ani-chiral phases. Here $\theta=0.32$. Right: Same as left but tilted. The plus/minus sign indicates the change in chirality hence existence of Hubert wall.}
%\label{vortex_wall_picture}
%\end{figure}
%%%%%%%%%%%%%%%%%%%%%%%%%%%%%%%%%%%%%%%%%%%%%%%%%%%%%%%%%%%%%%%%%%%%%%%%%%%%%%%%%%%%%%%%%%%%%%%%%%%%

\begin{figure}[t]
\includegraphics[width=8cm]{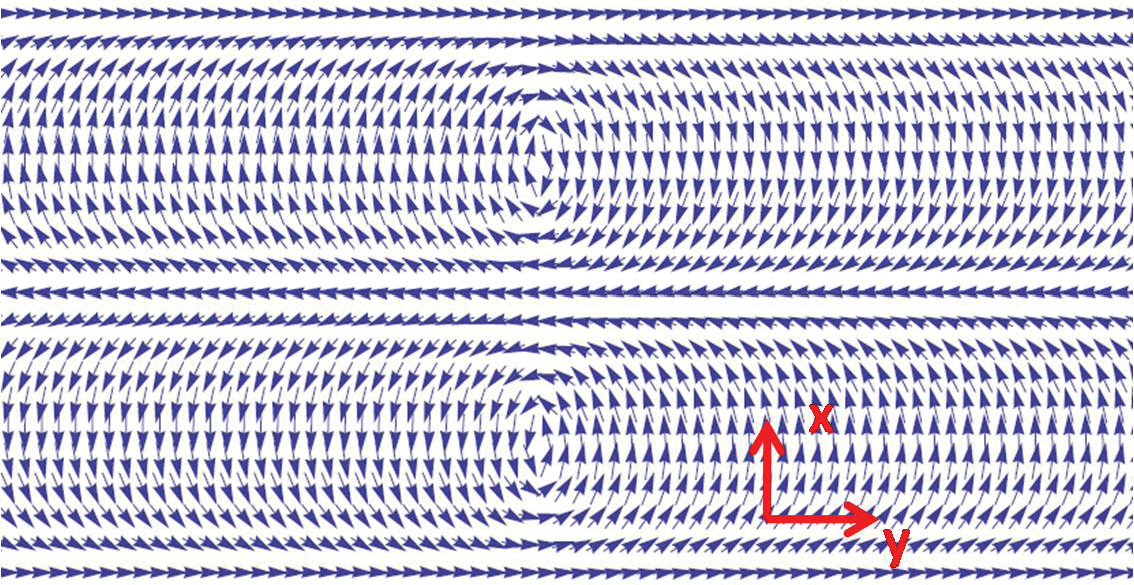}
\caption{Vector field illustration of numerically optimized interface between two
phases with opposite chirality. Here $\theta=0.32$. Spins are drawn to be in x-y plane while in the text they are assumed to be in y-z plane.}
\end{figure}

%%%%%%%%%%%%%%%%%%%%%%%%%%%%%%%%%%%%%%%%%%%%%%%%%%%%%%%%%%%%%%%%%%%%%%%%%%%%%%%%%%%%%%%%%%%%%%%%%%%%%%
\begin{figure}
\includegraphics[width=8cm]{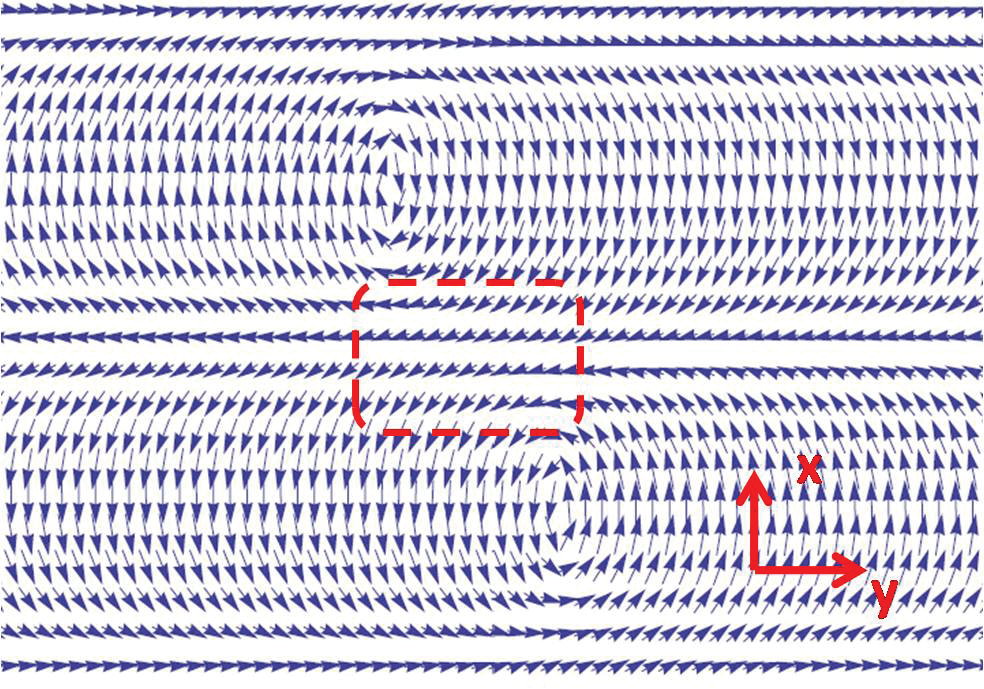}
\caption{Left: Vector field illustration of numerically optimized interface between two
 phases with opposite chirality for $\theta=0.32$. The vortices have moved away from each other along y direction. The dashed path closes around a Hubert wall.}
\label{vortex_wall_picture}
\end{figure}

%%%%%%%%%%%%%%%%%%%%%%%%%%%%%%%%%% ansatz verification %%%%%%%%%%%%%%%%%%%%%%%%%%%%%%%
\begin{figure}[t]
\includegraphics[width=8.5cm]{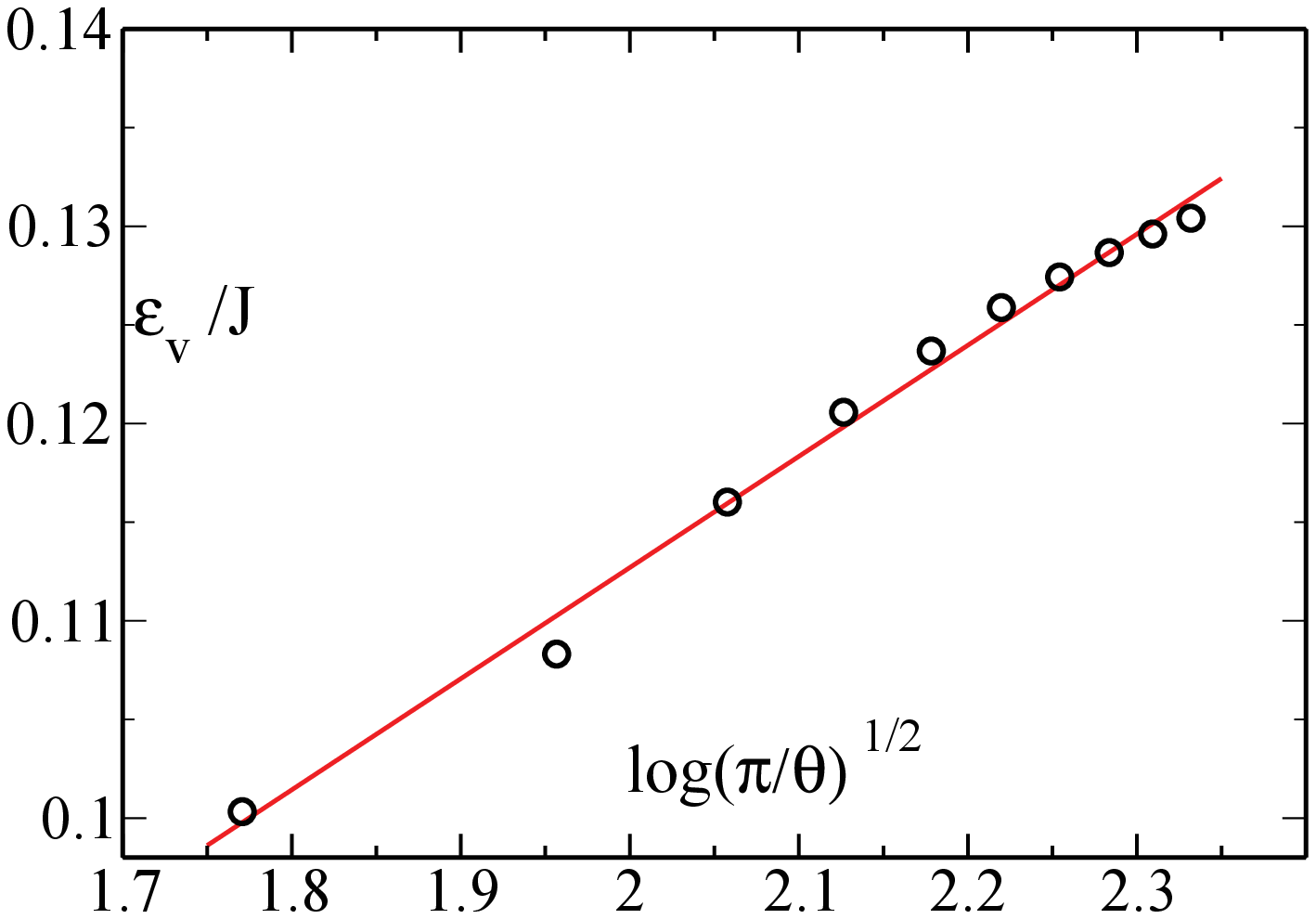}
\caption{Numerical comparison of the energy of the vortex domain wall per vortex(circles) with the
results of the ansatz(solid line). The numerical result is obtained by optimization of the discrete
model.}
\label{vortex_energy}
\end{figure}
%%%%%%%%%%%%%%%%%%%%%%%%%%%%%%%%%%%%%%%%%%%%%%%%%%%%%%%%%%%%%%%%%%%%%%%%%%%%%%%%%%%%%%%%%%%%%%%

%%%%%%%%%%%%%%%%%%%%%%%%%%%%%%%%%% ansatz verification %%%%%%%%%%%%%%%%%%%%%%%%%%%%%%%
\begin{figure}[t]
\includegraphics[width=8.5cm]{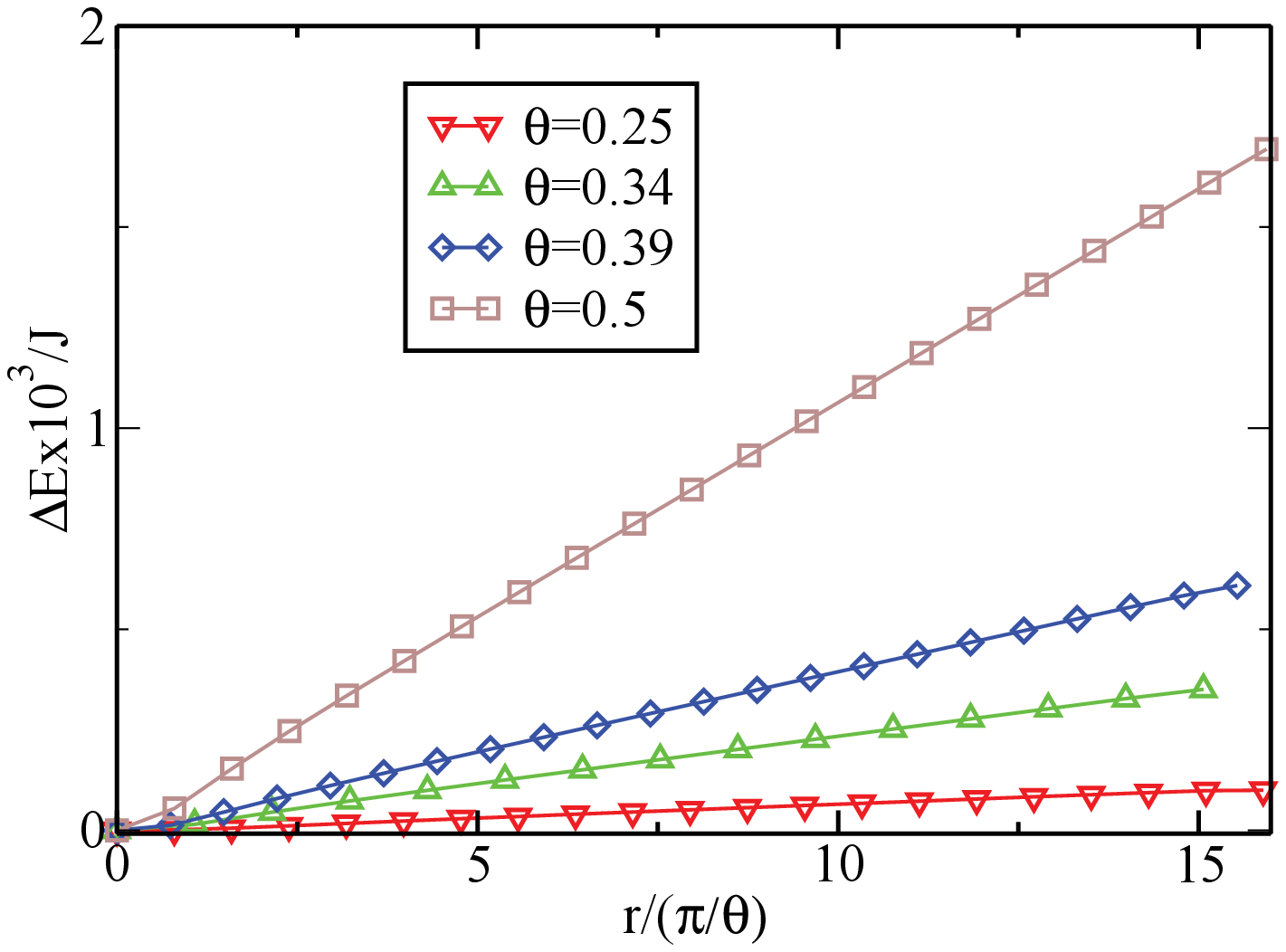}
\caption{Vortex interaction energy obtained as the difference $\Delta E$ between the energy of a flat wall and the energy of a flat wall with one vortex displaced by a distance $r$. The units are $10^{-3}J$ and $\pi a/\theta$ for energy and distance respectively. The units are so that the lattice constant is $a=1$. The results are plotted for few $\theta$'s.}
\label{Evplot}
\end{figure}
%%%%%%%%%%%%%%%%%%%%%%%%%%%%%%%%%%%%%%%%%%%%%%%%%%%%%%%%%%%%%%%%%%%%%%%%%%%%%%%%%%%%%%%%%%%%%%%

For a more realistic investigation of the properties of vortex domain walls in helical magnets it is
possible to use optimization techniques to find the lowest energy possible configurations of the
discrete model (\ref{discrete}) at zero temperature. It is also possible to investigate the interaction between vortices in such
configurations. In this section we show that the ansatz solution for the vortices
obtained previously are in good agreement with optimized results in this section. Also we show
that the interaction between vortices in a pure vortex wall are comparatively weak in agreement with our previous analytical speculations.
 This will justify the vicinal surface theory\cite{Nattermann} for the tilt of vortex domain walls via generation of staircases of Hubert walls and vortex lines. We will explain this theory in detail in the next section.

We use a two dimensional square lattice with the assumption that the system is symmetric in the third
direction i.e. the direction of vortex lines. We also use the limit of strong anisotropy and assume the vectors defined on each lattice points lie in the plane parallel to the helical axis. We start with a configuration consisting of two
helical domains with opposite chirality with a sharp interface between them and find the lowest
energy possible configuration using local iterative optimization methods. We use
periodic boundary conditions in the direction of helical axis thus the number of the vortices in the
wall is always even. For a wall with $n$ vortices(see Fig.\ref{vortex_wall_picture}) we find the total energy per unit of the length of
the wall, $\ell_n=n\pi/\theta$ (in units of lattice constant and $J=1$) obeys the following form:
\eqa
\varepsilon_n=\alpha+{\beta\over\ell_n}\sqrt{\log\ell_n}
\eqe
for $n>0$. Note that $n$ is also half the number of pitches. In our optimization $n=0$ does not apply because there needs to be at least one half pitch in the configuration. The constant $\alpha$ is the residual energy per unit of the length of the wall that comes from the two finite size helical structures on the two sides. Fitting gives $\beta\approx 3.54$. This gives the energy of one vortex to be proportional to
$(\log\pi/\theta)^{1/2}$ apart from an unimportant constant in agreement with the ansatz (Fig.\ref{vortex_energy}).

In order to make an estimate of the order of the interaction between these vortices in the wall we
take the numerically obtained solution and create a new configuration by artificially displacing one vortex out of the wall and use it as a starting point to obtain the
minimum solution again by the same optimization technique. Note that our optimization technique is able to find only the local minima and so at its best it finds the lowest energy possible for a wall with a displaced vortex. The fact that the energy of the solution reduces (Fig.\ref{Evplot}) as we repeat our procedure for smaller displacements from the wall indicates the stability of the wall. On the other hand it is interesting to see that this reduction in energy is very small ($\sim 0.001 J$). This is a strong evidence that the
vortices in the vortex wall almost do not interact. As explained in the previous section the small vortex-vortex interaction is because the helical background screens the vortex field.

\begin{figure}[h]
\includegraphics[width=5 cm]{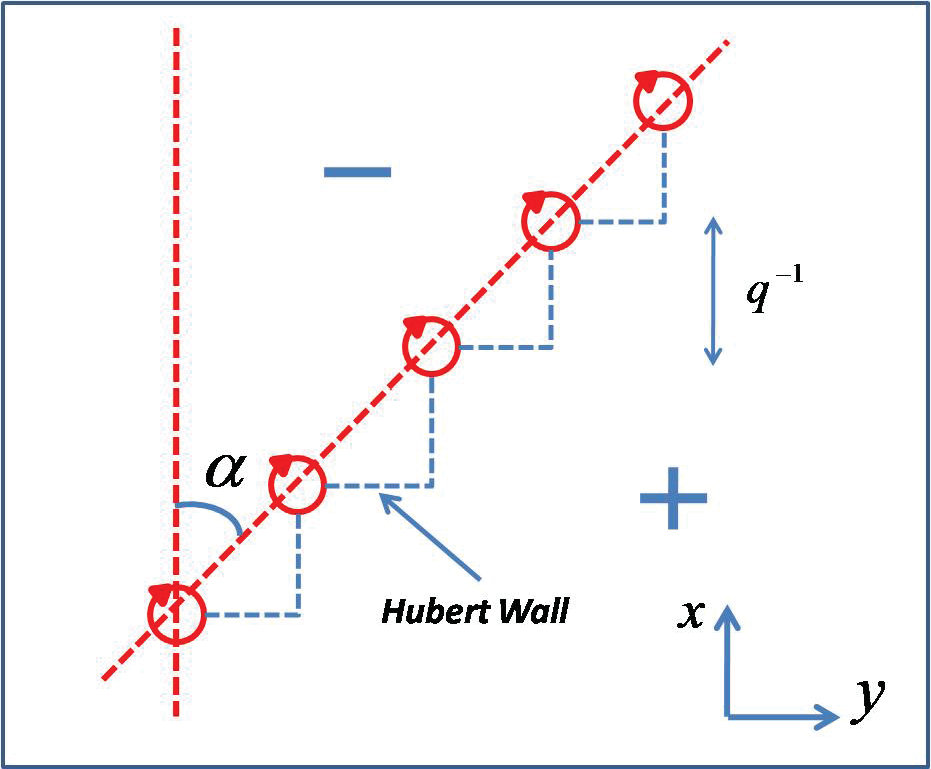}
\caption{Illustration of tilting of a vortex wall with movement of vortices along with generation of Hubert wall segments. Plus/minus sign indicate the chirality of the spiral domains. The red circles indicate the center of each vortex.}
\label{tilted_wall}
\end{figure}

The lowest energy fluctuations of such wall based on the above results then are the displacements of the vortices away from planar configuration which generates
%formation of staircases\cite{Nattermann}. These staircases are consisting of vortex lines displaced away from the array with a Hubert wall in between neighboring ones.
%It is easy to see this structure formation by considering a tilted array (see figure \ref{vortex_wall_picture} right).
Hubert walls. It is easy to see this structure formation by considering a tilted array (see Fig. \ref{vortex_wall_picture} right and also Fig. \ref{tilted_wall} for illustration). Between any two vortices that are displaced with respect to each other in the direction perpendicular to the wall there must be a Hubert wall formed. In the next section we derive a local elasticity Hamiltonian for such fluctuations.

%%%%%%%%%%%%%%%%%%%%%%%%%%%%%%%%%%%%%%%%%%%%%%%%%%%%%%%%%%%%%%%%%%%%%%%%%%%%%%%%%%%%%%%%%%%%%%%
\section{Tilted Domain Walls: Staircase formation}
\begin{figure}[h]
\includegraphics[width=8 cm]{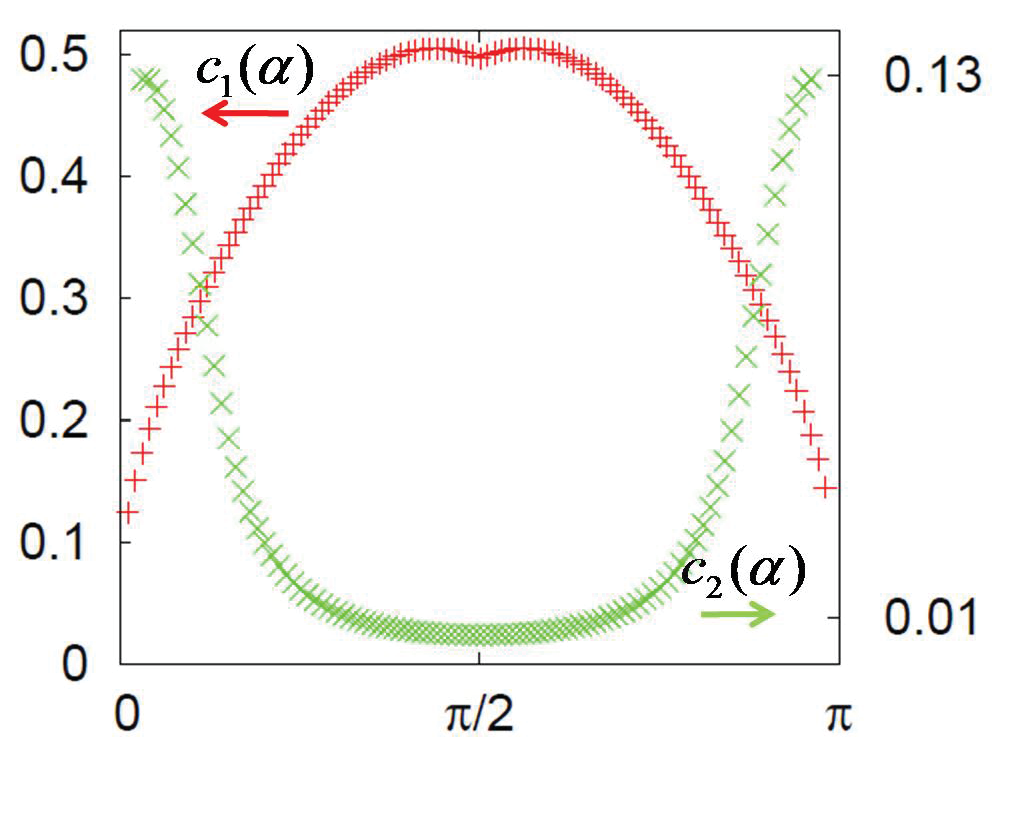}
\caption{Elastic constant of the vortex wall as a result of vicinal fluctuations as a function of orientation for $\theta=0.5$.}
\label{c1c2}
\end{figure}

We found out in sections \ref{numerics} and \ref{vortex wall} that the vortex lines in a vortex wall interact weakly. The fluctuations of the vortex wall are
basically displacements of the vortices with generation of Hubert segments (see Fig.\ref{tilted_wall}).
%A tilt of the vortex wall must then generate a staircase of vortex lines of width $\pi/q$. Recently a theory for elasticity of vortex
%walls has been developed\cite{Nattermann} based on \textit{vicinal surface} approach. In this approach the elasticity of the vortex wall is found out to be local. Here for the sake
%of completeness we explain in more detail the derivation:
Let's assume the tilt is around an axis parallel to ferromagnetic planes. This means the generated vortex
lines are parallel to each other. The density of vortex lines for such a tilt with an angle
$\alpha$ is $n(\alpha)=(q/\pi)\cos\alpha$. It is also important to remember that the energy of
vortex lines in the system is distributed highly anisotropic (Eq. \ref{vortex energy analytic}) so the energy of each
vortex line would depend on $\alpha$. As a result by adding the energies for the Hubert segments and
the vortex lines (neglecting the vortex interactions) the energy density per unit of the area of the wall will be:
\eqa\label{sigma_alpha}
\sigma(\alpha)&=& {\epsilon_v(\alpha)n(\alpha)+\sigma_H\over \left(1+{n(\alpha)^2\pi^2\over
q^2}\right)^{1/2}}\\
&=& \sin\alpha\left({q\over\pi}\epsilon_v(\alpha)\cot\alpha+\sigma_H\right)
\eqe
where $\sigma_H$ is the elastic constant of the Hubert wall introduced earlier. A small tilt $\delta\alpha$ away from this configuration
will cost an extra energy density to create vortex lines:
\eqa
\label{delta-sigma}
\Delta \sigma(\alpha,\delta\alpha) &=& {
\sigma(\alpha+\delta\alpha)L(\alpha+\delta\alpha)-\sigma(\alpha)L(\alpha)\over L(\alpha)}\nonumber\\
&\approx & \sigma'(\alpha)\delta\alpha+{1\over 2}[\sigma(\alpha)+\sigma''(\alpha)]\delta\alpha^2
\eqe
where $L(\alpha)$ is the linear size of the wall. We use this calculation for a local area around every point $\bx_\parallel$ on the wall. The elastic energy of this wall then would be:
\eqa
{\cal H}_e(\alpha)=\int d^2\bx_\parallel \Delta\sigma(\alpha,\delta\alpha(\bx_\parallel))
\eqe
Here $\bx_\parallel$ is the local coordinate of the points on the wall in continuous approximation. Note that the linear term in (\ref{delta-sigma}) is zero at finite temperatures because the vortices slide against each other almost freely so the average $\delta\alpha$ will be zero. In a long wavelength approximation the parameter $\delta\alpha(\bx_\parallel)\approx \partial_x u_y$ is the average wall distortion. With this
fact we conclude that the total energy of the staircase is:
\eqa\label{elastic_local}
{\cal H}_e(\alpha)=\int d^2\br_\parallel \left[c_1(\alpha)(\partial_x u)^2+c_2(\alpha)(\partial_z
u)^2\right]
\eqe
in which $c_1(\alpha)= 1/2[\sigma(\alpha)+\sigma''(\alpha)]$ and $c_2(\alpha)=\sigma(\alpha)$. We see that the above energy is local. Note that the term $(\partial_z u)^2 $ exists because of the
area change of the wall upon tilting around x-axis. This kind of tilt changes the length of the vortex lines in the wall and hence changes the total energy.
The above elastic energy resembles the simplest local quadratic distortion energy of conventional Block domain walls with a difference that the elastic
constants depend on the local coordinates. In a global tilt of the wall elastic constants depend on the tilting angle. These elastic constants can be numerically
calculated as a function of the tilt angle using equations (\ref{sigma_alpha}), (\ref{vortex energy analytic}) and definition of $\lambda$ presented in section \ref{vortex wall}.
The results are shown in Fig. \ref{c1c2}. The elastic constant $c_1(\alpha)$ represents the creation of staircases. Here $\alpha$ is the tilt of the wall around the $\hat{\bf z}$ direction. For $\alpha\approx \pi/2$ around the $\hat{\bf z}$ the elastic energy per unit area is lower because the vortex density is the lowest. For $\alpha\approx 0$ the vortex density is highest possible value which results in $c_1$ to have higher values. The constant $c_2(\alpha)$ on the other hand represents the energy cost for tilt around the $\hat{\bf x}$ direction. Small tilts ($\alpha\approx 0$) in this direction will only change the length of the vortex lines without creation of Hubert wall segments while for $\alpha\approx \pi/2$ the vortex wall disappears hence the elastic constant has the lowest value. On the other hand the cusp represents the creation/destruction of vortices.
%%%%%%%%%%%%%%%%%%%%%%%%%%%%%%%%%% pinning threshold ratio %%%%%%%%%%%%%%%%%%%%%%%%%%%%%%%
%%%%%%%%%%%%%%%%%%%%%%%%%%%%%%%%%%%%%%%%%%%%%%%%%%%%%%%%%%%%%%%%%%%%%%%%%%%%%%%%%%%%%%%%%%%%%%%
\section{Disorder}
We consider a major type of disorder in this paper namely defects in crystals where
magnetic atoms are missing on random sites. These defects are called vacancies. Usually domain walls gain exchange
energy by adopting to vacancy sites. In this case we attempt to derive a Hamiltonian for these types of disorder in continuum limit.
Domain walls experience friction force due to these types of disorder. External driving forces at zero temperature can mobilize these walls provided they are
stronger than a threshold value called depinning threshold. Periodic nature of the vortex walls however can have interesting effects in presence of disorder and driving force. The internal dynamics of vortex array helps avoiding the pinning centers hence lower dissipation. This effect can be partially captured in perturbation theory by observation that the \textit{effective mobility increases in the moving phase compared with the uniform domain walls}. Threshold values for both vortex wall and Hubert wall have been obtained elsewhere\cite{Nattermann} using the scaling relations on the same line of argument as we present here. We use perturbation theory in disorder strength for weak disorder. The range of the validity of our approach is for high enough velocities (see below) in which the width of the fluctuations of the vortex lines in the array is less than the array spacing $(\pi/q)$. This is similar to the analysis of the dynamics of vortex line lattices in Type-II superconductors\cite{Nattermann2000}.
\subsection{Dynamics of discrete domain walls}
In order to find the correction to mobility in the moving phase of the vortex wall we start with the overdamped
equation of motion for each vortex line in the array. Obviously each vortex line can move transverse to the wall in y-direction (indicated by $\perp$ sign ) and in longitudinal (x) direction, parallel to the domain wall (indicated by $\parallel$ sign). The equation of motion will then look like the following:
\eqa\label{dynamic-vec-eq}
\bmu^{-1}\cdot{\partial\bu_n\over\partial t}=-{\delta{\cal H}\over \delta \bu_n}+\bff(\bx,\bu_n)+{\bf h}
\eqe
in which $\bu_n=\{u_n(z),v_n(z)\}$ is the displacement vector of the vortex line number $n$ with $y$ and $x$ components and:
\eqa\label{friction}
\bff(\bx,\bu_n)=-{\delta{\cal H}_I\over \delta \bu_n}
\eqe
is the friction force density as a result of disorder represented by the Hamiltonian ${\cal H}_I$. Here ${\bf x}=\{x,z\}$ represents a point on the wall. The matrix $\bmu$ is the mobility matrix of the wall. This parameter determines the ratio of the overdamped velocity to the external force at different length scales. Also in the above ${\cal H}$ is the elastic Hamiltonian of the wall determined by equation (\ref{elastic_local}) however throughout the rest of this calculation we assume the most general elastic Hamiltonian:
\eqa\label{general_elastic}
{\cal H}=\int_\bk \left[\hat{\cal G}_\perp(\bk) |\hat {u}_y(\bk)|^2 + \hat{\cal G}_\parallel(\bk) |\hat {u}_x(\bk)|^2\right]
\eqe
in which $\hat{}$ indicates the Fourier transformation. The functions ${\cal G}_\perp$ and ${\cal G}_\parallel$ represent the elasticity of the wall in two different directions. In a conventional case these functions are quadratic functions of $\bk$. Finally ${\bf h}$ is the deriving force density presumably applied uniformly across the wall. Note that the equation of motion (\ref{dynamic-vec-eq}) is only valid in the limit of overdamped motion of vortices when the damping forces are dominant. The dynamics of the magnetic moments is sufficiently explained by Ginzburg-Landau-Gilbert equation. In such analysis the out of plane component of the moments $m_x$ (recall that vorticity is in y-z plane) included in the vortex gyrovector\cite{Tatara2008} must be included to obtain the equation of motion but it does not change the dynamics as long as it is almost constant in time. Based on this fact our overdamped approximation is valid in which the Gilbert damping is represented by the mobility $\bmu$\cite{Tatara2008}.

 To determine the disorder energy, ${\cal H}_I$ we use the fact that the wall consists of a one dimensional periodic array of vortex lines:
\begin{equation}\label{general disorder}
 {\cal H}_I=\int_\br {\cal K}(\br){\cal E}(\br)
\end{equation}
in which we have summed up the energy gain contribution of each vortex from vacancy sites:
\eqa
{\cal E}(\br)=\sum_{n=-N_v}^{N_v} \epsilon_v[x-n\pi/q-v_n(z),y-u_n(z)]
\eqe
and $\epsilon_v(x,y)$ is the energy density of a vortex line at point ${\bf r}=\{x,y,z\}$. The function ${\cal K}(\br)$ determines the effect of disorder:
\eqa
{\cal K}(\br)=\gamma\sum_{i=1}^{N_I}\left[f_I(\br-\br_i)-n_I\right]
\eqe
with $f_I(\br)$ being the vacancy/impurity form factor $\int_\br f_I(\br)=1$ and $n_I$ is their
density. Throughout this article we assume the form factor vanishes in a range of the order of the crystal lattice constant. The sum in the above is over the position of impurities/vacancies $\br_i$ which is random, $\gamma$ is the effective volume occupied by impurity/vacancy  , $N_I$ is the number of impurity sites and $n_I=N_I/\Omega$ the impurity concentration. The above energy can be written as:
\begin{eqnarray}\label{pinning-H}
{\cal H}_I &=& \int_\br{\cal K}(\br)\sum_n \epsilon_v[x-n\pi/q-v_n(z),y-u_n(z)]=\nonumber\\
%&=& \int_{\br,\br'} {\cal K}(\br)\epsilon_v(x-x',y-y')\rho({\bf r}')\nonumber\\
&=& \int_{\br} V(\br)\rho_v({\bf r}).
\end{eqnarray}
where we have defined:
\begin{eqnarray}\label{density+potential}
 V({\bf r}) &=& \int_{\br'} {\cal K}(\br')\epsilon_v(x-x',y-y')\delta(z-z') \\
 \rho_v({\bf r}) &=& \sum_n \delta[y-u_n(z)]\delta[x-n\pi/q-v_n(z)]
\end{eqnarray}
as impurity potential and vortex line density. Here we approximate the vortex line density as if the displacement of vortex lines vary smoothly over the domain wall as a result we can treat the displacement vector as a smooth function of $x$: $\bu_n(z)\approx \bu(x,z)$. The impurity energy will be then of the following form:
\eqa
{\cal H}_I=\int d^{2}\bx \ V(\bx,u_y)\rho(x-u_x)
\eqe
where we have defined $\rho(x)=\sum_n \delta(x-n\pi/q)$. The impurity potential in the above is dependent on the position distribution of impurities which is assumed to be completely random. The spatial average and correlation of this random potential can be calculated:
\eqa
\left\langle V(\br)\right\rangle &=& 0 \nonumber\\
\Delta(\br) = \left\langle V(\br)V({\bf 0})\right\rangle &=& \gamma^2n_I f_v(\br)
\eqe
where:
\eqa
\langle A(\br) \rangle=\int \Pi_{i=1}^{N_I} {d^3r_i\over \Omega} A(\br,\br_i)
\eqe
means averaging over disorder and:
\eqa
f_v(\br)=\delta(z)\int dx'dy' \epsilon_v(x-x',y-y')\epsilon_v(x',y')
\eqe
is the average vortex correlation energy at different positions in the medium. This function is the maximum in the range of the core size of the vortices $(\approx q^{-1})$ and vanishes quickly over larger scales.

 The force density vector (\ref{friction}) has components perpendicular and parallel to the
domain wall plane $\bff=\{f_\parallel,f_\perp\}$:
\eqa\label{forces}
f_\perp(\bx,\bu) &=& -{\partial V\over\partial u_y}\rho(x-u_x) \\
f_\parallel(\bx,\bu) &=& V(\bx,u_y)\rho'(x-u_x)
\eqe
%the equation \ref{dynamic-vec-eq} becomes:
%\eqa
%\bmu^{-1}\cdot{\partial\bu\over\partial t}=-\int d^D\bx_\parallel'{\boldsymbol{\cal
%G}}^{-1}(\bx_\parallel-\bx_\parallel')\cdot \bu(\bx_\parallel') +\bff(\bx,\bu)+{\bf h}
%\eqe
Application of the perturbation in powers of disorder will only give us the force-velocity
relation of the system at high velocities. While this approach will not ultimately lead to a
relation in the full range of external force magnitudes it provides us with an estimate for the
effective mobility of the system and also it will ultimately help us to estimate the threshold
force density in large length scales in a renormalization group point of view\cite{Nattermann1992}(see the appendix for a full discussion). Let's now use the
perturbation theory order by order to clarify the above argument: First we introduce the average drift
velocity of the vortex array $\bv$ which in our case has only a non-zero component perpendicular to the
domain wall plane $\bv = v\bhy$. The remaining displacements of the vortex lines then can be separated:
\eqa
\bu(\bx,t)={\bv} t +\bxi(\bx,t)
\eqe
There is of course no reason that all the vortex lines in the array have the same drift velocity in a random environment. On the other hand the array is the boundary of a domain which is expanding or shrinking hence all the vortex lines will need to drift together on average. This is why in a frame of reference moving with velocity $\bv$ the vortex array is not seen to have uniform array structure but looks \textit{corrugated} instead:
\eqa
\langle \bxi(\bx,t)\rangle = {\bf g}(\bx)
\eqe
This effect does not exist in more conventional Neel or Bloch type domain walls but here the internal degree of freedom of the wall dictates such changes. This effect however is conventional in studies of pinned two and three dimensional vortex line lattices\cite{Nattermann2004} in type II superconductors. We will present the resulting corrugation later in this section.

Now we expand the out of plane displacement in orders of disorder strength:
\eqa
\bxi=\bxi^{(0)}+\bxi^{(1)}+\bxi^{(2)}+\ldots
\eqe
after averaging the equation of motion for the zeroth and first order we will have:
\eqa
\bmu^{-1}\cdot \bv &=& \bh \\
\bxi^{(0)} &=& 0
\eqe
%and:
%\eqa
%\bmu^{-1}\cdot {\partial\bxi^{(1)}\over\partial t} &=& -\int_{\bx'} {\boldsymbol{\cal
%G}}^{-1}(\bx-\bx')\cdot\bxi^{(1)}(\bx',t)+\nonumber\\
%&+& \bff(\bx,\bv t)
%\eqe
%after averaging:
%\eqa
%\bmu^{-1}\cdot\bv &=& \bh \\
%{\bf g}^{(1)} &=& 0.
%\eqe
which is nothing but the force-velocity relation of a rigid wall in disorder free medium. For the first order approximation:
\eqa
\hat{\xi}_y^{(1)}(\bk,\omega) &=&\int_{Q}{ -i{\omega\over v^2}\hat{V}(k_x+Q,{\omega\over
v},k_z)\hat{\rho}(-Q)\over i\mu_\perp^{-1}\omega+{\hat{\cal G}}_\perp^{-1}(\bk)}\\
\hat{\xi}_x^{(1)}(\bk,\omega) &=&\int_{Q}{ -i{Q\over v}\hat{V}(k_x+Q,{\omega\over
v},k_z)\hat{\rho}(-Q)\over i\mu_\parallel^{-1}\omega+{\hat{\cal G}}_\parallel^{-1}(\bk)}
\eqe
in which we have assumed the elastic energy for fluctuations parallel and transverse to the domain wall are governed by two distinct functions ${\cal G}_\parallel$ and ${\cal G}_\perp$ in the form of equation \ref{general_elastic}. In the above, in the limit of uniform wall ($\hat{\rho}(Q)\propto \delta_{Q,0}$) the resulting displacements approach to the conventional Block wall results\cite{Hubert}. For the longitudinal displacement of the vortex lines obviously $\hat{u}_x\rightarrow 0$ in the limit of uniform wall. From this result we obtain for the width of the longitudinal (internal) fluctuations of the vortex array the follwoing:
\begin{widetext}
\eqa
W^2(\bx-\bx',t-t')=\left\langle [\xi_x(\bx,t)-\xi_x(\bx',t')]^2\right\rangle \approx {q^2 \hat{\Delta}(0)\over v\sigma_\parallel^2}\left[1-e^{-q|z-z'|}\cos q(x-x')\right]\delta_{\omega_0}(t-t')
\eqe
\end{widetext}
The above result has been obtained by considering that the correlation of disorder $\Delta(\br)$ is short range with the range of the order of $q^{-1}$. Also we have assumed quadratic elasticity $\hat{\cal G}(\bk)=\sigma_\parallel k^2$. Of course $W(\bx,t)$ is periodic in direction of vortex array $\hat{\bx}$ but the width of the fluctuations of each vortex, $W(0,0)$ vanishes quickly on length scales larger than $q^{-1}$. The function $\delta_{\omega_0}(x)$ is a smooth delta function which vanishes on frequency scales shorter than $\omega_0\approx \mu_\parallel\sigma_\parallel q^2$. The associated time scale $\omega_0^{-1}$ is the time scale corresponding with the reaction of internal vortex wall degrees of freedom to friction forces. We will find this scale in a more intuitive analysis later in the discussion section. From this result we can now find out the range of the validity of the perturbation approach by requiring $W<\pi/q$ which sets the limit of $v> v_i=\left(\omega_0 q^4\hat{\Delta}(0)/\pi^2\sigma_\parallel^2\right)$. This result indicates the range at which we can treat the vortex wall dynamics perturbatively and at the same time \textit{do not ignore its internal degrees of freedom}. Using previous approximations\cite{Li2012} $\sigma_\parallel\approx J q/a$ we can estimate the value of $v_i\approx \gamma^2 n_I\omega_0|e_v/J|^2|aq/\pi|^2$ in which $e_v$ is the vortex line energy in the wall per unit length. For vortex arrays with low density of vortex lines without the screening of the vortex interaction the average vortex energy grows as the size of the system which sets a very unrealistic limit for the value of $v_i$ however for vortex walls in helical magnets the vorticity is screened away from the core of the vortex and the average energy per unit area of the vortex wall is $(\approx (Jq/a)\log aq)$ which will result in $v_i\approx 51.2\times joules\times m^{-3}\times \mu_\parallel $ for typical values of disorder concentration $n_I\approx 10^{19} cm^{-3}$ in Holmium.

 For second order approximation in disorder on the other hand:
\eqa
\bff(\bx,\bu)\approx\bff(\bx,\bv t)+\bxi\cdot\nabla_\bu\bff(\bx,\bv t).
\eqe
We insert this into the equation of motion and average over disorder. This will result in:
\begin{widetext}
\eqa\label{2nd order}
\mu_\perp^{-1} v-h=-\int_{\bx'}{\cal
G}_\perp^{-1}(\bx-\bx')g_\perp^{(2)}(\bx')+\left\langle\bxi^{(1)}\cdot\nabla_\bu f_\perp(\bx,\bv
t)\right\rangle
\eqe
\end{widetext}
in which we have introduced $g_\perp^{(2)}$ as the y-component of the function ${\bf g}$ in the second order of disorder. After averaging, the second term in the right hand side will only depend on $x$ through the function $\rho(x)$ (see Eq. \ref{forces}) which is periodic. Because the left hand side is independent of $x$ and $z$ the first term on the right hand side must only depend on $x$ as well. Consequently $g_\perp^{(2)}$ must only
depend on $x$ and also must be periodic. This equation is presenting the first correction to the effective mobility of an overdamped vortex wall motion. Even more than that this equation determines the structural changes of the moving vortex array caused by external force and friction forces. Taking Fourier transform and integrating over one
period using the fact that $\hat{g}_\perp^{(2)}(0)=0$:
\begin{widetext}
\begin{equation}\label{v-h-2}
{\pi\over q}(\mu_\perp^{-1}v-h) =\int_{\bk,\bQ}\left[{iQ_y^3\over
\hg_\perp(\bk)-i\mu_\perp^{-1}Q_yv}+{-iQ_yQ_x^2\over
\hg_\parallel(\bk)-i\mu_\parallel^{-1}Q_yv}\right]|\hat{\rho}(Q_x)|^2\hat{\Delta}(\bk,Q_y)
\end{equation}
\end{widetext}
Using the real part of the above equation one can find a correction to mobility at non-zero velocity:
\eqa
v &=& \tilde{\mu}_\perp h \\
\tilde{\mu}_\perp &=& {\mu_\perp\over 1-{q\over\pi}\ell_D}
\eqe
in which:
\eqa
\ell_D &=& \ell_\perp + \ell_\parallel \nonumber\\
\ell_\perp &=& -\int_{\bk,\bQ} {Q_y^4|\hat{\rho}(Q_x)|^2\hat\Delta(k_x+Q_x,k_z,Q_y)\over
\left[\hg_\perp(\bk)\right]^2+(\mu_\perp^{-1}v Q_y)^2}\\
\ell_\parallel &=& {\mu_\perp\over\mu_\parallel}\int_{\bk,\bQ}
{Q_y^2Q_x^2|\hat{\rho}(Q_x)|^2\hat\Delta(k_x+Q_x,k_z,Q_y)\over
\left[\hg_\parallel(\bk)\right]^2+(\mu_\parallel^{-1}v Q_y)^2}\nonumber\\
\eqe
Note that the imaginary part of equation (\ref{v-h-2}) is zero because it is an integration over odd derivatives of $\Delta(\br)$ which is assumed to be an even function.

 Using the fact that $\hat{\Delta}(\bk,Q)=\gamma^2 n_I |\hat{\epsilon}_v(k_x,Q)|^2$ ($\hat{\epsilon}_v$ does not depend on $k_z$) is always positive one can see that $\ell_\perp <0$
corresponding to the out of plane fluctuations of the wall trying to reduce the effective mobility
of the wall while $\ell_\parallel >0$ tries to increase it, an effect which depends on the
gradient of the density of the wall (note the term $|Q_x\hat{\rho}(Q_x)|^2$ in the second equation). Also  $\ell_\parallel=0$ for uniform density. This is
physically plausible since the lattice structure of the wall while driven, helps the wall to avoid resisting
centers during its motion. From an energetics point of view the kinetic energy injected into the array by the external force now has another channel to be stored in instead of being dissipated by the friction. This will reduce the dissipation hence increase the mobility. The length $\ell_D$ can be interpreted as the average amount of displacement of each vortex line while interacting with impurity centers. While $\ell_\perp$ is positive the $\ell_\parallel$ coming from internal degrees of freedom tries to reduce it.

 Let's assume a simple quadratic form for the elasticity of the wall: $\hat{\cal G}_\parallel(\bk)=\sigma_\parallel k^2$ and $\hat{\cal G}_\perp(\bk)=\sigma_\perp k^2$. Also we assume short range behavior of $\Delta(\br)$ and isotropic mobility $(\mu=\mu_\parallel=\mu_\perp)$ and elastic constants $(\sigma=\sigma_\parallel=\sigma_\perp)$ for both directions perpendicular and parallel to domain wall plane we can approximate the above integrals as:
\eqa
\label{ell}
\ell_\perp &\approx &- {q^4\hat{\Delta}(0)\over \sigma^2}{\cal F}_\perp\left({v\over v_p}\right) \\
\ell_\parallel &\approx & {q^4\hat{\Delta}(0)\over \sigma^2}{\cal F}_\parallel\left({v\over v_p}\right) \\
\eqe
Here we have assumed simple quadratic elasticity: $\hat{\cal G}_\perp=\hat{\cal G}_\parallel=\sigma k^2$. The functions ${\cal F}_\perp(x)$ and ${\cal F}_\parallel(x)$ must in principle be calculated numerically however by adopting an exponential model form for the function $\Delta(\br)$ they will be of the form:
\eqa
{\cal F}_\perp(x) &=& \int d^3p  {p_y^4 e^{-(p_x^2+p_y^2)}\over (p_x^2+p_z^2)^2+x^2 p_y^2} \\
{\cal F}_\parallel(x) &=& \int d^3p  {p_y^2 e^{-(p_x^2+p_y^2)}\over (p_x^2+p_z^2)^2+x^2 p_y^2} \\
\eqe
The velocity $v_p=\mu\sigma q$ is the characteristic velocity lower than which the effects of internal degrees of freedom of the vortex wall appear. We will obtain these parameters later in discussion section by a more intuitive argument.
\subsection{Corrugation of The Vortex Wall}
In this section we briefly analyze the structural properties of the domain wall in presence of disorder.
Our starting point is the equation (\ref{2nd order}) from which we can deduce the average displacement of the vortex wall at higher velocities.
Since the second term on the right hand side of this equation is periodic the function $g_\perp^{(2)}(x)$ will have to be periodic as well. After some algebra
one obtains for $g_\perp^{(2)}(x)=\sum_m g_m e^{imqx}$ the following:
%\begin{widetext}
\eqa
g_m &=& {c_m\over m^2}(1-\delta_{m,0})+\sum_{n\neq 0}\left[ {c_{m+n}\over m^2}+q^2 {n(n+m)\over m^2}d_m\right]\nonumber\\
\eqe
%\end{widetext}
in which:
\eqa
c_m={\mu_\perp \over\sigma_\perp\pi^2}\int_\bk {-ik_y^3\hat{\Delta}(\bk)\over \sigma_\perp\mu_\perp\left[k_z^2+(mq-k_x)^2\right]-ik_y v} \nonumber\\
d_m={\mu_\parallel q^2\over\pi^2\sigma_\perp}\int_\bk {-ik_y\hat{\Delta}(\bk)\over \sigma_\parallel\mu_\parallel\left[k_z^2+(mq-k_x)^2\right]-ik_y v} \nonumber\\
\eqe
Interestingly we see that the above coefficients are all zero for $v=0$ which means the corrugations appear or enhance when the wall starts to move. It is also possible to estimate the value of the above integrals at the moving phase of the wall:
\eqa
c_{\pm 1}\approx d_{\pm 1} &\approx & n_I\gamma^2\left(J\mu\over a^5 v\right)^{1/2} q
\eqe
where we have the same simplifications as previous section for mobility and elastic constants. Also in the above we have used a cutoff for the integral over $k_x$ to include only wavelength of the order of $q^{-1}$. Finally smaller wavelengths ($|m|> 1$) in the corrugation have been neglected for large scale approximations.
\section{Discussion and Conclusions}
In this paper we have presented an analytic model as well as numerical verification of the predicted vortex walls in both clean and disordered helical magnets. From an analytic point of view we showed how the helical background screens the vorticity effect in long distances resulting in the weak interaction between vortex lines. We have also presented a detailed discussion of the dynamics of such wall in the presence of disorder. In this calculation we have also shown that a moving vortex wall array will display on average \textit{spatial corrugations}. In this discussion we have also revealed how the periodic nature of the wall will result in correction of the mobility toward higher values compared with other conventional magnetic domain boundaries. Here we can further discuss such effect by time scale considerations. The time scale during which a domain wall with a thickness of $\sim q^{-1}$ passes a pinning center during its motion is of the order of $\tau_\perp\sim q^{-1}/v$. On the other hand the effect of this impact throughout the wall propagates during a time scale $\tau_\parallel\sim \lambda^2/\sigma_\parallel\mu_\parallel$ in which $\lambda$ is the length of propagation. The ratio of the two time scales then will be: $\tau_\parallel/\tau_\perp \sim \theta(\lambda/\xi_\parallel)^2$ in which $\xi_\parallel=(\sigma_\parallel\mu_\parallel a/v)^{1/2}$ is the typical correlation length between fluctuations throughout the wall. In order for the periodic nature of the wall to have any effect in the dynamics of the wall the propagation length must be at least of the order of the period of the vortex line array $\lambda\sim q^{-1}$. These considerations result in a range of velocities $v\leq v_p\sim \sigma_\parallel\mu_\parallel q=\omega_0 /q$ at which the periodic effect show up. The velocity $v_p$ appeared already in approximating the internal distorsions of the vortex lines which try to increase the mobility of the wall in equation (\ref{ell}). The time scale $\omega_0^{-1}$ here is again appearing because we are estimating the response time of the internal structure of the wall to external perturbations. The limiting velocity $v_p$ will then directly determine the range of external force densities $h\leq h_p=v_p/\mu_\perp\sim \sigma_\parallel q \mu_\parallel/\mu_\perp$ at which the periodic structure affects the motion of the wall. This value is estimated to be $h_p\approx 6.6\times 10^{11} joules/m^3$ using typical values of Holmium parameters\cite{Jensen}. It is interesting to see that $v_i/v_p\approx 10^{-11}$ for Holmium indicates an accessible range of force densities/velocities at which the internal fluctuations of the vortex wall can be effective in its dynamics.
\begin{acknowledgments}
  I would like to thank Prof. Thomas Nattermann for his critical supervision and Institute for Theoretical Physics at the University of Cologne in Germany for hosting most of this work. This work was supported in the most part by German Sonderforschungsbereich 608 and also by US NSF-DMR 1054020. Also I would like to thank Prof. Yogesh N. Joglekar and the Indiana University-Purdue University at Indianapolis for their support.
\end{acknowledgments}
\section{Appendix}

 The depinning threshold naturally follows from setting $v\rightarrow 0$ in (\ref{v-h-2}):
\begin{widetext}
\eqa
{\pi\over q}h_c &=&\int_{\bk,\bQ}\left[-iQ_y^3{\hat{\cal G}}_\perp(\bk)+iQ_yQ_x^2{\hat{\cal
G}}_\parallel(\bk)\right]|\hat{\rho} (Q_x)|^2\hat {\Delta}(k_x-Q_x,k_z,Q_y)\\
&=& \int_{\bx} \left[{\cal G}_\perp(\bx)\Delta^{(3)}(\bx,0)C(x)+{\cal
G}_\parallel(\bx)\Delta^{(1)}(\bx,0)D(x)\right]
\eqe
\end{widetext}
in which:
\eqa
C(x) &=& \int_{x'}\rho(x')\rho(x-x')\\
D(x) &=& \int_{x'} \rho'(x')\rho'(x'-x)
\eqe
and the derivatives of the correlator are introduced as follows:
\eqa
\Delta^{(1)}(\br) &=& \langle \partial_y V(\br) V({\bf 0})\rangle \\
\Delta^{(3)}(\br) &=& \langle \partial_y^2V(\br)\partial_y V({\bf 0})\rangle
\eqe
Here by analysing the above result we try to explain in details an approximate method used before\cite{Nattermann} to estimate the threshold force density. The
first assumption is that the correlator of disorder potential $\Delta(\br)$ and its
spatial derivatives are of short range with an average value of $\Delta_L^{(n)}$ (for $n$th
derivative) at a length scale $L$. The length scale dependence comes from a renormalization point
of view where one systematically sums up the contributions from smaller scales into the correlator
using methods such as perturbation etc.\cite{Nattermann1992} With this assumption and taking into account the fact that
$D(x),C(x)>0$ we arrive at the estimate for the threshold of force density at the length scale $L$:
\eqa\label{threshold}
h_c\approx \left[{\cal G}_\perp(L)\Delta_L^{(3)}+{\cal
G}_\parallel(L)\Delta_L^{(1)}\right]
\eqe
On the other hand for an isotropically distributed set of pinning centers, $\Delta^{(n)}$ is generically
zero for odd $n$ resulting in the zero threshold force density however it is well known\cite{Nattermann1992} that in a
renormalization group flow $\Delta_L^{(4)}$ develops a pole at scales larger than Larkin
length scale $\cal L$ where the energy gain due to collective weak disorder fluctuations can
overcome the elastic deformation energy cost. This means $\Delta_L^{(2)}(\bx,u)$ at $L\approx {\cal L}$ has a
cusp near $\{\bx,u\}=0$ resulting in $\Delta_L^{(3)}$ being nonzero. The second term from the
parallel (longitudinal) deformations however remains zero at all length scales. Using this relation then
one can estimate the threshold force density without having to worry about the
physical value of the mobility. This will give us the maximum value of the threshold force density possible because for scales
higher than the Larkin length the disorder energy gain tends to decrease although it stays
higher than deformation energy cost.

The above argument also leads us to a way to determine the Larkin length scale of the problem.
Here we ignore the in-plane fluctuations of the wall and treat the vortex wall as a wall with
uniform density. This will still give us the upper bound for the threshold force density because
the internal fluctuations of the wall tends to reduce the magnitude of the threshold force as
we will see at the end of this section. We start with the one-loop renormalization group flow
equation for the force correlator $R(\bx,u)=\langle
f_\perp(\bx,u)f_\perp({\bf 0},0)\rangle=-\Delta^{(2)}(\bx,u)$. Provided we are close to the upper
critical dimension we can use the renormalization group equation for $R_L$\cite{Nattermann1992}:
\eqa
{dR_L(u)\over d L} &=& {\cal A}(L){d^2\over
du^2}\left[R_L(u)(2R_L(0)-R_L(u))\right]\label{rg1}\nonumber\\
{\cal A}(L)\delta L &=& \int_{1/L}^{1/(L-\delta L)} {d^2k\over 8\pi^2}\hat{\cal G}^2(\bk)
\eqe
taking two times derivative of the first equation in (\ref{rg1}) and setting $u=0$ and assuming
$R_L^{(3)}(0)=0$ as an analytic even function, we find the following solution up to the zeroth order
in $\epsilon$:
\eqa
[R_a^{(2)}(0)]^{-1}-[R_L^{(2)}(0)]^{-1}=-6\int_{1/a}^{1/L} {d^2k\over 8\pi^2}{\hat{\cal G}}^2(\bk)
\eqe
in which $R_a(u)$ is at $L=a$ the bare correlation function. As was mentioned earlier this
solution is divergent at the Larkin length given by the following equation at zeroth order of
$\epsilon$:
\eqa\label{larkin}
R_a^{(2)}(0)=\left[6\int_{1/a}^{1/{\cal L}} {d^2k\over 8\pi^2}{\hat{\cal G}}^2(\bk)\right]^{-1}
\eqe
In order to find the renormalized non-zero value of $R'(0)=\Delta^{(3)}(0)$ at Larkin length scale
we need to use the functional renormalization group equation again. Assuming ${\cal A}(L)=A_0
L^{\gamma-1}$ we use the following ansatz:
\eqa
R_L(u)=\alpha  L^{-a} R^\star(\beta uL^{-\zeta})
\eqe
The parameter $\alpha$ is fixed by the condition that the fixed point function $R^\star(0)=1$.
Inserting this into the RG equation results in $a=-2\zeta+\gamma$ and:
\eqa
R_L'(0)=L^{-\gamma/2}[R_L(0)]^{1/2}\left(-2\zeta+\gamma\over 2A_0\right)^{1/2}
\eqe
On the other hand using the fact that $\int_{-\infty}^{+\infty} R_L(x)dx$ is invariant under the
flow we obtain the exponent $\zeta$ with the physical assumption that $R^\star(x)\rightarrow 0$ as
$x\rightarrow \pm\infty$ to be $\zeta=\gamma/3$.

\end{document}